\documentclass[]{aastex61}
\usepackage{amsmath}
\submitjournal{ApJ}

\shorttitle{IY Lyr}
\shortauthors{Li et al.}
\begin{document}

\title{IY Lyr: A Thick-Disk first-overtone RR Lyrae Star with a Possible Neutron Star Companion}

\correspondingauthor{Linjia Li, Liying Zhu}
\email{lipk@ynao.ac.cn, zhuly@ynao.ac.cn}

\author{Linjia Li}
\affiliation{Yunnan Observatories, Chinese Academy of Sciences,
P.O. Box 110, Kunming 650216, People's Republic of China}
%\affiliation{Key laboratory of the structure and evolution
%of celestial objects, Chinese Academy of Sciences, P.O. Box 110, Kunming,
%650216, People's Republic China}

\author{Shengbang Qian}
%\affiliation{Yunnan Observatories, Chinese Academy of Sciences, P.O. Box 110, Kunming, 650216, People's Republic China}
\affiliation{Department of Astronomy, School of Physics and Astronomy, Yunnan University, Kunming 650091, People's Republic of China}
\affiliation{Key Laboratory of Astroparticle Physics of Yunnan Province, Yunnan University, Kunming 650091, People's Republic of China}
%\affiliation{Key laboratory of the structure and evolution
%of celestial objects, Chinese Academy of Sciences, P.O. Box 110, Kunming,
%650216, People's Republic China}
%\affiliation{University of Chinese Academy of Sciences, No.1 Yanqihu East Rd, Huairou District, Beijing, 101408, People's Republic China}

\author{Ildar Asfandiyarov}
\affiliation{Ulugh Beg Astronomical Institute, Uzbekistan Academy of Sciences,
33 Astronomicheskaya Street, Tashkent 100052, Uzbekistan}

\author{Azizbek Matekov}
\affiliation{Yunnan Observatories, Chinese Academy of Sciences,
P.O. Box 110, Kunming 650216, People's Republic of China}
%\affiliation{Key laboratory of the structure and evolution
%of celestial objects, Chinese Academy of Sciences, P.O. Box 110, Kunming,
%650216, People's Republic China}
\affiliation{Ulugh Beg Astronomical Institute, Uzbekistan Academy of Sciences,
33 Astronomicheskaya Street, Tashkent 100052, Uzbekistan}
\affiliation{University of Chinese Academy of Sciences, No.1 Yanqihu East Rd, Huairou District, Beijing 101408, People's Republic of China}

\author{Liying Zhu}
\affiliation{Yunnan Observatories, Chinese Academy of Sciences,
P.O. Box 110, Kunming 650216, People's Republic of China}
%\affiliation{Key laboratory of the structure and evolution
%of celestial objects, Chinese Academy of Sciences, P.O. Box 110, Kunming,
%650216, People's Republic China}
\affiliation{University of Chinese Academy of Sciences, No.1 Yanqihu East Rd, Huairou District, Beijing 101408, People's Republic of China}

\author{Boonrucksar Soonthornthum}
\affiliation{National Astronomical Research Institute of Thailand, 191 Siriphanich Building, Huay Kaew Road, Chiang Mai 50200, Thailand}

\author{Evelina Gaynullina}
\affiliation{Ulugh Beg Astronomical Institute, Uzbekistan Academy of Sciences,
33 Astronomicheskaya Street, Tashkent 100052, Uzbekistan}

\author{Alina Khalikova}
\affiliation{Ulugh Beg Astronomical Institute, Uzbekistan Academy of Sciences,
33 Astronomicheskaya Street, Tashkent 100052, Uzbekistan}

\author{Jiajia He}
\affiliation{Yunnan Observatories, Chinese Academy of Sciences,
P.O. Box 110, Kunming 650216, People's Republic of China}

\author{Fangbin Meng}
\affiliation{Yunnan Observatories, Chinese Academy of Sciences,
P.O. Box 110, Kunming 650216, People's Republic of China}
%\affiliation{Key laboratory of the structure and evolution
%of celestial objects, Chinese Academy of Sciences, P.O. Box 110, Kunming,
%650216, People's Republic China}
\affiliation{University of Chinese Academy of Sciences, No.1 Yanqihu East Rd, Huairou District, Beijing 101408, People's Republic of China}

\author{Huiting Zhang}
%\affiliation{Yunnan Observatories, Chinese Academy of Sciences, P.O. Box 110, Kunming, 650216, People's Republic China}
\affiliation{Department of Astronomy, School of Physics and Astronomy, Yunnan University, Kunming 650091, People's Republic of China}
\affiliation{Key Laboratory of Astroparticle Physics of Yunnan Province, Yunnan University, Kunming 650091, People's Republic of China}
%\affiliation{Key laboratory of the structure and evolution
%of celestial objects, Chinese Academy of Sciences, P.O. Box 110, Kunming,
%650216, People's Republic China}
%\affiliation{University of Chinese Academy of Sciences, No.1 Yanqihu East Rd, Huairou District, Beijing, 101408, People's Republic China}

\author{Jiangjiao Wang}
\affiliation{Yunnan Observatories, Chinese Academy of Sciences,
P.O. Box 110, Kunming 650216, People's Republic of China}
%\affiliation{Key laboratory of the structure and evolution
%of celestial objects, Chinese Academy of Sciences, P.O. Box 110, Kunming,
%650216, People's Republic China}
\affiliation{University of Chinese Academy of Sciences, No.1 Yanqihu East Rd, Huairou District, Beijing 101408, People's Republic of China}

\author{Xiangdong Shi}
\affiliation{Yunnan Observatories, Chinese Academy of Sciences,
P.O. Box 110, Kunming 650216, People's Republic of China}
%\affiliation{Key laboratory of the structure and evolution
%of celestial objects, Chinese Academy of Sciences, P.O. Box 110, Kunming,
%650216, People's Republic China}



\begin{abstract}

IY Lyr, historically misclassified as an eclipsing binary, has been previously identified as a first-overtone RR Lyrae star (RRc star). Using multiband photometry (All-Sky Automated Survey for Supernovae, Zwicky Transient Facility, TESS, and our $BVRI$ data), Large Sky Area Multi-Object Fiber Spectroscopic Telescope spectroscopy, and Gaia astrometry, we investigate its pulsation, binarity, and Galactic population. From O--C analysis, we detect a long-term period decrease and a light-travel time effect with an orbital period of 3.94 $\pm$ 0.09 years, eccentricity of 0.46 $\pm$ 0.15, and a mass function of 0.65 $\pm$ 0.14 M$_{\odot}$. The companion is independently supported by radial velocity residuals and Gaia proper motions. Combined constraints yield an orbital inclination of 94.2$^{\circ}$ $\pm$ 1.1$^{\circ}$ and a companion mass of 1.37 $\pm$ 0.19 M$_{\odot}$. Chemical abundances ([Fe/H] $\simeq$ -1.0 $\pm$ 0.1, [$\alpha$/Fe] $\simeq$ +0.27 $\pm$ 0.03, Xiang et al. 2019) and dynamics ($L_{\rm z}$ $\simeq$ 1287 $\pm$ 35 kpc km s$^{-1}$, $Z_{\rm max}$ $\simeq$ 1.17 $\pm$ 0.10 kpc) identify IY Lyr as likely an old, high-$\alpha$, thick-disk star. The companion mass lies at the peak of the neutron star mass distribution, and the system's age excludes a main-sequence star; we conclude the companion is most likely a typical neutron star, although a massive white dwarf near the Chandrasekhar limit cannot be ruled out. IY Lyr is among the few RRc binaries with a compact companion supported by multiple methods, and it has important implications for thick-disk binary evolution and neutron star formation.

\end{abstract}




\section{Introduction} \label{sec:intro}

RR Lyrae stars (RRLs) are low-mass pulsating variables located on the horizontal branch, characterized by helium-burning cores and atmospheres that pulsate via the $\kappa$-mechanism \citep{2004rrls.book.....S,2015pust.book.....C}. They are primarily classified into ab-type RRLs (RRab stars), which exhibit large-amplitude fundamental-mode pulsations with sawtooth-shaped light curves in the optical, and c-type RRLs (RRc stars), which display smaller-amplitude first-overtone pulsations with sinusoidal light curves. Rare d-type RRLs (RRd stars) show dual-mode pulsations (\citealt{2021MNRAS.507..781N} and references therein). Studies of their pulsation reveal complex phenomena like the century-old Blazhko effect, a periodic modulation potentially linked to mode coupling, magnetic activity, or resonance \citep{2011rrls.conf..100K}; and the Oosterhoff dichotomy, which offers insights into the Galactic assembly history (refer to \citealt{2025MNRAS.542.1791L} and references therein). Their period-luminosity relation makes them essential standard candles for measuring distances, playing a crucial role in both stellar astrophysics and the study of galactic evolution \citep{2004rrls.book.....S}.

Despite the well-established prevalence of binary systems among stars, confirmed binary RRLs and robust candidates remain exceptionally rare (e.g., \citealt{2024A&A...691A.108S} and references therein). Detecting companions is crucial not only for understanding the binary evolution pathways that lead to old horizontal branch stars but also for improving the precision of RRLs as distance indicators, since binarity can influence their observed properties and period-luminosity calibration. The most widely used detection method exploits the star's own pulsation as a clock, searching for the light-travel time effect (LTTE) caused by orbital motion through analysis of timing residuals in Observed-minus-Calculated (O--C) diagrams \citep{2005ASPC..335....3S}. However, This signal can be degenerate with intrinsic period variations typical of these stars \citep{2025A&A...697A.154B}. Direct detection via eclipses in light curves is highly unlikely due to the stars' evolutionary history; having previously ascended the red giant branch, they likely possess widely separated companions with long orbital periods\footnote{\url{http://spiff.rit.edu/richmond/asras/rrlyr/rrlyr.html}}. To date, only one convincing eclipsing binary system containing an RRL has been identified (OGLE-BLG-RRLYR-02792, \citealt{2012Natur.484...75P,2013MNRAS.428.3034S}). Complementary techniques include radial velocity (RV) monitoring, which is observationally intensive due to the need for dense phase coverage, and astrometric analysis using high-precision data from missions like Gaia to detect companion-induced wobble in proper motion \citep{2019A&A...623A..72K,2019A&A...623A.116K,2022A&A...657A...7K}. It is anticipated that the release of the next Gaia dataset will yield more results in this area\footnote{\url{https://www.cosmos.esa.int/web/gaia/dr4}}. Nonetheless, in the era of large photometric surveys, the O--C method remains the primary discovery tool, with other techniques playing vital roles in confirmation and characterization.

Most RRL binary candidates discovered through the O--C method are RRab stars \citep{2015MNRAS.449L.113H,2016MNRAS.459.4360L,2018ApJ...863..151L,2019MNRAS.487L...1P,2021ApJ...915...50H}, while samples of first-overtone RRc stars remain rare. Among the few RRc binary candidates reported in the literature, two notable examples illustrate the challenges of confirmation. \citet{2004MNRAS.354..821D} reported the discovery of the RRc star BE Dor (MACHO* J050918.712-695015.31), which exhibited rapid period changes and suggested the LTTE in a binary system as a possible explanation. However, follow-up spectroscopic observations found no supporting RV variations, effectively ruling out the binary hypothesis \citep{2021tsc2.confE..53D}. This conclusion was later reinforced by an independent analysis of its complex period variations \citep{2022MNRAS.510.6050L}. A similar case is KIC 2831097, where significant phase variations were initially attributed to binary motion involving an $\geq$ 8.4 M$_{\odot}$ companion \citep{2017MNRAS.465L...1S}, but subsequent detailed RV monitoring failed to confirm binarity \citep{2025A&A...703A.286P}.

On the theoretical modeling front, \citet{2024MNRAS.52712196B} demonstrated through detailed binary evolution models that metal-rich RRLs may predominantly originate from binary interactions. In such systems, a companion star strips the envelope from a red giant branch primary, causing it to become bluer on the horizontal branch and enter the pulsation instability strip. This evolutionary channel naturally accounts for the existence of young, metal-rich RRLs, which are challenging to explain through single-star evolution, and motivates ongoing searches for such systems \citep{2025A&A...695L..14A}.

\begin{figure*}
\centering
%\vbox to4.0in{\rule{0pt}{5.5in}}
%\special{psfile=Figure01.eps angle=0 hoffset=0 voffset=0 vscale=100
%hscale=100}
\includegraphics[width=0.45\textwidth]{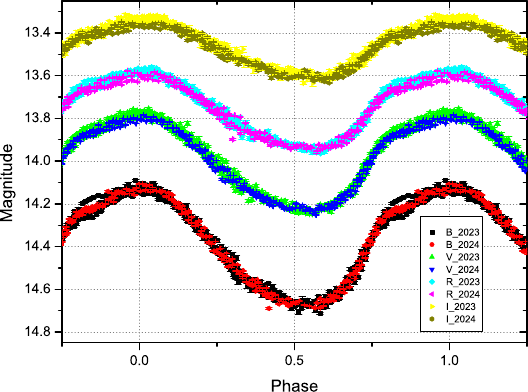}
\caption{Phased-folded $BVRI$ light curves of IY Lyr obtained with the ZEISS-600 telescope located at the Maidanak Astronomical Observatory. Observations were carried out during two runs: 2023 September--October and 2024 July--September. Data from the two runs are plotted with different symbols and colors as indicated in the legend. The phases are calculated by the linear ephemeris HJD$_{\rm MAX}$ = 2460209.2725775 + 0$^{\rm d}$.32661702 $\cdot$ $E$.} \label{Fig.1}
\end{figure*}

Against this backdrop, the star IY Lyr presents an intriguing case. It was historically misclassified as a W Ursae Majoris-type eclipsing binary with a 0.6531646 day period \citep{1951VeSon...1..407H}, an interpretation followed in several subsequent studies \citep{2006A&A...446..785M,2008OEJV...94....1B,2014AJ....147..119C,2020RAA....20..163Q}. However, its true nature as a pulsating RRc variable is supported by multiple recent surveys and catalogs \citep{2014AJ....148..121K,2017AJ....153..204S,2018AJ....156..241H,2019A&A...622A..60C,2020ApJS..249...18C,2022ApJ...931..131M,2023A&A...674A..18C}. The 0.653 day period reported in earlier statistical catalogs (e.g., \citealt{2020RAA....20..163Q}) simply reflects the historical misclassification. In contrast, the RRc classification is firmly established by the light-curve morphology, period, and amplitude shown in Figure \ref{Fig.1}. For instance, the Gaia DR3 RR Lyrae catalog identifies IY Lyr as an RRc star with a period of 0.32661639 day and a $G$ band amplitude of 0.3772 mag \citep{2023A&A...674A..18C}. Despite this clarification, a dedicated study of IY Lyr, particularly regarding a potential binary companion, has been lacking. Therefore, we monitored it using a 60 cm telescope in Uzbekistan and collected multi-epoch data from several survey projects (see Section \ref{sec:ObsAndSurv}). Our investigation reveals compelling evidence for a compact companion of approximately 1.4 $M_{\odot}$. Furthermore, we find IY Lyr to be metal-rich and a member of the thick-disk population characteristics that provide important context for its evolutionary status. We present our data analysis in Section \ref{Sec:Analysis}, followed by a discussion and conclusions in Sections \ref{Sec:Discussion} and \ref{Sec:Conclusion}.

\section{Observation and surveys} \label{sec:ObsAndSurv}

\subsection{Photometric data} \label{sec:Sect2.1}

Observations of IY Lyr were conducted over 15 nights between 2023 September-October and 2024 July-September using the ZEISS-600 telescope at the Maidanak Astronomical Observatory. This telescope is a 0.6 m Cassegrain system with a focal ratio of f/12.5, making it well suited for high-resolution stellar observations. It is equipped with a scientific-grade FLI IMG ProLine $1024\times1024$ CCD camera, which provides precise, low-noise imaging. Standard Johnson-Cousin Bessel $BVRI$ filters were used throughout the observing runs. The comparison star was NOMAD-1-1209-0294636 (R.A. = 18$^{\rm h}$:29$^{\rm m}$:47.48$^{\rm s}$, Decl. = +30$^{\circ}$:59$'$:12.57$''$), whose $BVRI$ magnitudes ($B$ = 14.2914, $V$ = 13.7291, $R$ = 13.3883, and $I$ = 13.0399) were derived from Gaia $G_{\rm BP}$, $G$, and $G_{\rm RP}$ photometry using the transformation relations from \citet{2025RASTI...4...37R}. The check star was NOMAD-1-1210-0294366 (R.A. = 18$^{\rm h}$:29$^{\rm m}$:50.37$^{\rm s}$, Decl. = +31$^{\circ}$:00$'$:48.00$''$), with magnitudes $B$ = 14.4192, $V$ = 13.8812, $R$ = 13.5537, and $I$ = 13.2115. Figure~\ref{Fig.1} presents the phase-folded light curves in the $BVRI$ bands.%The color index curves (B-V, V-R, V-I) shown in Figure 2 were constructed by dividing the pulsation phase interval [0,1] into multiple equal bins. Within each bin, magnitudes from each filter were averaged, with errors propagated accordingly. Color indices were then obtained by subtracting the averaged magnitudes of two different bands at the same phase bin.

To fully utilize multi-epoch photometric data for the study of IY Lyr, this work combines publicly available data from two major sky surveys: the Zwicky Transient Facility (ZTF) and the All-Sky Automated Survey for Supernovae (ASAS-SN). ZTF employs a wide-field camera at Palomar Observatory to conduct high-cadence photometric surveys, offering high sampling rates and broad sky coverage, which are advantageous for time-domain astronomy \citep{2019PASP..131a8003M}. ASAS-SN, through a global network of small-aperture telescopes, provides long-term, continuous monitoring of bright targets across the entire sky, supplying essential data for analyzing long-term photometric behavior \citep{2014ApJ...788...48S}. Table~\ref{Table1} summarizes the key information of the photometric data used in this study, including the observation time range, corresponding passband, number of valid data points, and data source, thereby clearly presenting the observational coverage and statistical characteristics of each dataset. Although the space telescope TESS does not provide ready-made light-curve products \citep{2015JATIS...1a4003R}, we followed the data reduction approach described in \citet{2023AJ....166...83L} and obtained approximately 32,000 data points. These data will be used for further analysis in Sections \ref{Sec:OCAnalysis} and \ref{sec:RadialVelocity}.

In addition, to precisely determine the pulsation phase of IY Lyr, we conducted an additional one-hour observation on the evening of 2025 November 25, using the 85 cm telescope at the Xinglong Observatory of the National Astronomical Observatories, Chinese Academy of Sciences. From these data, a set of times of maximum light was derived for subsequent O--C analysis. The Xinglong photometry was obtained contemporaneously with spectroscopy from the same site to enable phase determination; only the photometric data are used in this work.

\subsection{Spectral information} \label{sec:Sect2.2}

The Large Sky Area Multi-Object Fiber Spectroscopic Telescope (LAMOST), a major Chinese scientific facility, has conducted medium- and low-resolution spectroscopic observations of IY Lyr \citep{2012RAA....12..723Z}. Based on these data, several studies have derived stellar physical parameters for this target, including RV, effective temperature, and metallicity \citep{2018yCat.5153....0L,2019yCat.5164....0L,2022yCat.5156....0L,2019ApJS..245...34X,2023ApJS..266...40W,2024ApJS..272...31W,2020ApJS..246....9Z,2022ApJS..258...26Z,2024ApJS..271...58D,2026yCat.5162....0L} . In this work, we primarily adopt the parameters provided by \citet{2024ApJS..272...31W} for further analysis, as their study specifically focuses on RR Lyrae variables and employs a targeted data-processing methodology, making their parameters particularly suitable for our investigation.

\subsection{Astrometric data} \label{sec:Sect2.3}

Astrometric data provide two-dimensional information about the projection of an object's orbit onto the celestial tangent plane. By combining proper motion data from the Hipparcos and Gaia satellites, \citet{2019A&A...623A..72K,2022A&A...657A...7K} developed and applied a method to detect close companions through proper motion anomalies. This approach was subsequently extended to search for companions of RR Lyrae variables \citep{2019A&A...623A.116K}. Inspired by this work, we have developed a new method that, starting from known orbital parameters (e.g., derived from LTTEs or RV measurements), uses differences in proper motion data from different epochs to solve for the remaining orbital parameters, such as the longitude of the ascending node ($\Omega$) and the orbital inclination ($i$). A detailed derivation of the method is presented in Section \ref{Sec:ProperMotions}.
Typically, this method requires combining proper motion data from Hipparcos and Gaia. However, IY Lyr is relatively faint and was not recorded in the Hipparcos catalog. We note that the proper motion values reported in Gaia DR2 and Gaia DR3 exhibit significant differences within their respective uncertainties (see Section \ref{Sec:ProperMotions} for details). Therefore, in this study, we attempt to use the proper motion differences between the Gaia data releases to constrain and confirm the orbital parameters of IY Lyr.

\section{Analysis} \label{Sec:Analysis}

In this section, we present a comprehensive analysis to detect and characterize the binary companion of IY Lyr. First, we extract the orbital parameters from the O--C diagram. Next, we independently verify these parameters using RV and astrometric proper motion data. Finally, we determine the system's kinematic properties and its membership within the Galactic population.

\subsection{O--C analysis} \label{Sec:OCAnalysis}

For pulsating variable stars, the times of light maximum are the most commonly used phase reference for O--C analysis. In this study, 180 times of light maximum for IY Lyr were determined based on multi-epoch photometric data. The following methods were employed to obtain accurate timings according to the characteristics of the different data sources.

1. \textit{Data from our own observations using the 60 cm telescope in Uzbekistan and the 85 cm telescope at Xinglong Station.} The conventional polynomial fitting method around the light-curve maximum was employed. Specifically, data points near the maximum were selected and fitted with an algebraic polynomial. The time of light maximum was then determined by locating the zero-point of the first derivative of the fitted curve \citep{2005ASPC..335....3S}. This approach is suitable for light curves with relatively high time resolution and dense phase coverage.

2. \textit{Survey data (ZTF, ASAS-SN, TESS).} Since these data typically cover complete pulsation cycles over extended time baselines, we fitted the entire phase-folded light curves using Fourier series and determined the times of maximum by locating the extrema of the fitted function. The Fourier polynomial is expressed as follows:
\begin{equation}
m(t)=A_{0}+\sum^{n}_{k=1}A_{k}\sin[\frac{2\pi k t}{P_{\rm pul}}+\phi^{\rm s}_{k}], \label{Equ:1}
\end{equation}
where $m(t)$ is the magnitude observed at time $t$, $A_{0}$ is the mean magnitude, $A_{k}$ and $\phi^{\rm s}_{k}$ are the amplitude and phase of the $k$-th component, respectively, and $P_{\rm pul}$ is the pulsation period. To account for different noise levels and sampling properties, the order $n$ was chosen as follows: $n=3$ for ZTF and ASAS-SN data, and $n=5$ for TESS data due to their higher quality and denser sampling. This approach follows the same strategy as described by \citet{2021AJ....161..193L,2022MNRAS.510.6050L}.

\begin{figure*}
\centering
%\vbox to4.0in{\rule{0pt}{5.5in}}
%\special{psfile=Figure01.eps angle=0 hoffset=0 voffset=0 vscale=100
%hscale=100}
\includegraphics[width=0.45\textwidth]{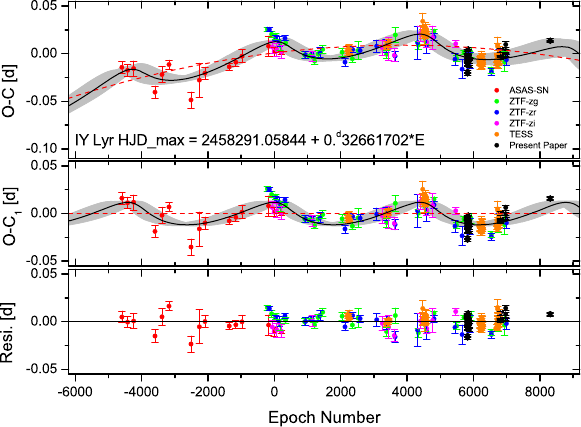}
\caption{Upper panel: The O--C diagram for IY Lyr. The black solid line refers to a combination of downward parabolic and periodic variations. The red dashed line represents the parabolic variation. The light gray shaded area represents the range within one standard deviation of the best-fit model. Middle panel: The O--C residuals from the quadratic term (O--C$_{1}$) to IY Lyr. The black solid line refers to the periodic variation. Bottom panel: The O--C residuals from the both the quadratic term and the periodic variation (Resi.) to IY Lyr.} \label{Fig.2}
\end{figure*}

All determined times of light maximum, along with their associated errors, observational methods, and data sources, are listed in Table~\ref{Table2}. For our own observed data, the error of the time of light maximum is obtained by multiplying the root-mean-square residual of the fit by the square root of the sum of squared partial derivatives of the extremum with respect to each observation. For survey project data, the errors of the times of light maximum are obtained based on Method I in \citet{2006Ap&SS.304..363M}. Using the corrected linear ephemeris:
\begin{equation}
\rm HJD_{\rm max} = 2458291.05844(163) + 0.^{d}32661702(33) \cdot \emph{E}, \label{Equ:2}
\end{equation}
originally comes from \citet{2018MNRAS.477.3145J}, we calculated the corresponding O--C values and plotted the O--C diagram (upper panel of Figure~\ref{Fig.2}). To describe the linear period change and the possible LTTE, we fitted the data using the following model:
\begin{equation}
O--C=\delta T_{0}+\delta P_{0}\cdot E+\frac{\beta}{2}E^{2}+\tau, \label{Equ:3}
\end{equation}
where $\delta T_{0}$ and $\delta P_{0}$ are the corrections to the initial epoch and the pulsation period, respectively; $\beta$ is the linear rate of period change (in day cycle$^{-1}$); and $\tau$ denotes the time delay caused by the LTTE. The expression for $\tau$ is
\begin{equation}
\tau=(\frac{a_1 \sin i}{c})[\sqrt{1-e^{2}}\sin E^{*}\cos\omega+\cos E^{*}\sin\omega], \label{Equ:4}
\end{equation}
where $a_{1} \sin i / c$ represents the projected semi-major axis in days, $e$ is the orbital eccentricity, $\omega$ is the longitude of periastron, and $E^{*}$ denotes the eccentric anomaly (for details, see \citealt{2014MNRAS.444..600L,2018ApJ...863..151L}). In the fitting process, all O--C points were assigned equal weight. This follows the common practice in O--C analyses of pulsating variables, where datasets often combine heterogeneous sources and the formal uncertainties do not fully capture intrinsic cycle-to-cycle scatter. Weighting by inverse variance would strongly bias the fit toward the most recent high-precision data, at the expense of the earlier points that define the long-term trend. The fitting results are presented in Table~\ref{Table3}. The parabolic component corresponds to a period change rate of $\beta$ = -1.24 $\pm$ 0.14 days Myr$^{-1}$, indicating that the pulsation period of IY Lyr is decreasing. After removing the parabolic trend, the O--C residuals exhibit cyclic variations (middle panel of Figure~\ref{Fig.2}), which can be explained by the LTTE. The derived orbital period is $P_{\rm orb}$ = 1438.5 $\pm$ 31.4 days (about 3.94 yr) with an eccentricity of $e$ = 0.46 $\pm$ 0.15.

\subsection{Orbital component in RV} \label{sec:RadialVelocity}

To independently verify the binary hypothesis, we analyze RV variations using spectroscopic observations of IY Lyr from the LAMOST survey. Two epochs are available: a low-resolution spectrum (LRS) obtained on UTC 2015-05-26T17:44:20.26, and a medium-resolution spectrum (MRS) taken on 2020-06-15T16:38:00.00. RV measurements from these spectra specifically for metal lines and the H$\alpha$ line are adopted from \citet{2024ApJS..272...31W}. These data contain blended signals, including the systemic velocity of the binary, pulsation-induced velocities, and orbital motion contributions. Our aim is to isolate the orbital component by subtracting the pulsation contribution.

We correct for pulsation-induced RV variations using published RV templates for RRLs \citep{2017ApJ...848...68S,2021ApJ...919...85B,2024RAA....24g5009H}. The corrected RVs are listed in Table~\ref{Table4}. Several key steps in this process are outlined below:

1. \textit{Determination of the Pulsation phase.} The observation time is obtained from the DATE-OBS keyword in the LAMOST FITS headers and converted to Heliocentric Julian Date (HJD). To compute the pulsation phase accurately, a linear ephemeris is necessary. Because pulsation modulation may occur, ephemerides can vary between observing epochs. Fortunately, contemporaneous photometry is available for both LAMOST epochs: ASAS-SN observations coincide with the LRS epoch, and TESS observations correspond to the MRS epoch. From these data, we derive the following linear ephemerides:
\begin{equation}
\rm HJD_{\rm max} = 2457184.78478 + 0.^{d}3266093 \cdot \emph{E}, \label{Equ:5}
\end{equation}
and
\begin{equation}
\rm HJD_{\rm max} = 2459014.84237 + 0.^{d}3265670 \cdot \emph{E}. \label{Equ:6}
\end{equation}

2. \textit{Pulsation RV amplitude.} The $g$-band light curve from ZTF shows a pulsation amplitude of 0.48 mag for IY Lyr. Using Equations (5) and (6) from \citet{2024RAA....24g5009H}, we convert this to RV amplitudes of 25.1 $\pm$ 2.1 km s$^{-1}$ for metal lines (Mg) and 35.6 $\pm$ 6.6 km s$^{-1}$ for the H$\alpha$ line. These amplitudes are then used to scale the template curves prior to subtraction.

3. \textit{Phase alignment of templates.} The zero-point of the \citet{2021ApJ...919...85B} template corresponds to the mean RV epoch, not to the light maximum. Therefore, we shift the template phase by +0.12 so that the minimum RV of the Mg template aligns with phase 0, as defined above.

4. \textit{Offset correction for template mean RV.} The mean RV of the \citet{2024RAA....24g5009H} templates is nonzero: -6.325 km s$^{-1}$ for the Mg lines and -9.415 km s$^{-1}$ for the H$\alpha$ line. We apply the corresponding offsets to the template values before performing the subtraction.

\begin{figure*}
\centering
%\vbox to4.0in{\rule{0pt}{5.5in}}
%\special{psfile=Figure01.eps angle=0 hoffset=0 voffset=0 vscale=100
%hscale=100}
\includegraphics[width=0.85\textwidth]{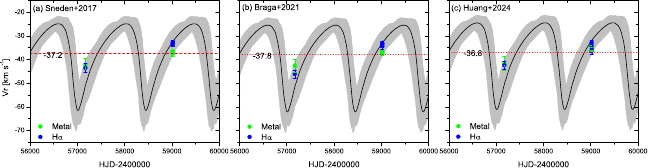}
\caption{Pulsation-corrected RVs of IY Lyr as a function of time. The corrections are applied using three different template sets: (a) \citet{2017ApJ...848...68S}, (b) \citet{2021ApJ...919...85B}, and (c) \citet{2024RAA....24g5009H}. The solid curves show the orbital RV variations predicted from the O--C solution. The light gray shaded area represents the range within one standard deviation of the best-fit model. The good agreement between the corrected velocities and the predicted curves independently supports the binary interpretation.} \label{Fig.3}
\end{figure*}

Figure~\ref{Fig.3} displays the RV residuals after pulsation subtraction, plotted against time. Green symbols represent RVs derived from the Mg lines, while blue symbols correspond to those from the H$\alpha$ lines. The solid black curve illustrates the orbital RV variation predicted from the O--C derived orbital elements; the gray shaded region indicates the 1$\sigma$ confidence interval obtained through error propagation. A horizontal dotted red line marks the mean systemic velocity, obtained by minimizing the sum of squared residuals between the observed data points and the theoretical orbital curve. The resulting systemic velocities from the different templates are -37.2, -37.8, and -36.8 km s$^{-1}$, with a mean value of -37.3 $\pm$ 3.0 km s$^{-1}$, which we adopt as the systemic RV of the IY Lyr binary system.

The agreement between the RV residuals and the predicted orbital curve supports the presence of a companion, corroborating the evidence from O--C analyses. This consistency across independent observational techniques reinforces the binary interpretation of IY Lyr.

\subsection{Orbital parameter constraints based on Gaia proper motions} \label{Sec:ProperMotions}

When a star exhibits orbital motion caused by a companion, its proper motion on the celestial sphere consists of two components: the systemic proper motion and the periodic orbital proper motion induced by the binary orbit. Inspired by methods that use proper motion anomalies to detect unseen stellar companions \citep{2019A&A...623A..72K,2019A&A...623A.116K,2022A&A...657A...7K}, we combine results from O--C analysis with proper motion data to further constrain the orbital parameters of the IY Lyr system.

The fitted parameters ($a_1 \sin i$, $e$, $\omega$, $P_{\rm orb}$, $T$) from the O--C analysis provide information on six orbital parameter elements. The quantity $a_1 \sin i$ combines two orbital parameters: the semi-major axis of the pulsating primary star's orbit, $a_{1}$, and the orbital inclination, $i$. $e$ is the eccentricity, $\omega$ the longitude of periastron, $P_{\rm orb}$ the orbital period, and $T$ the time of periastron passage. The seventh parameter, the longitude of the ascending node $\Omega$, remains undetermined. To obtain the complete set of Keplerian orbital elements ($a_{1}$, $i$, $e$, $\omega$, $P_{\rm orb}$, $T$, and $\Omega$), additional positional or proper motion data on the celestial sphere are required.

A Cartesian coordinate system is defined on the celestial sphere, with the x-axis pointing north along the decl. direction and the y-axis pointing east along the R.A. direction. The target's orbital motion, described by the seven parameters mentioned above, yields the components of the orbital position:
\begin{equation}
x_{\rm orb}=A(\cos E^{*}-e)+F \sqrt{1-e^{2}} \sin E^{*}, \label{Equ:7}
\end{equation}
and
\begin{equation}
y_{\rm orb}=B(\cos E^{*}-e)+G \sqrt{1-e^{2}} \sin E^{*}, \label{Equ:8}
\end{equation}
where $E^{*}$ is the eccentric anomaly as mentioned above, and $A$, $B$, $F$, and $G$ are the Thiele--Innes constants:
\begin{equation}
A=a_{1}[\cos\Omega\cos\omega - \sin\Omega\sin\omega\cos i], \label{Equ:9}
\end{equation}
\begin{equation}
B=a_{1}[\sin\Omega\cos\omega + \cos\Omega\sin\omega\cos i], \label{Equ:10}
\end{equation}
\begin{equation}
F=a_{1}[-\cos\Omega\sin\omega - \sin\Omega\cos\omega\cos i], \label{Equ:11}
\end{equation}
\begin{equation}
G=a_{1}[-\sin\Omega\sin\omega + \cos\Omega\cos\omega\cos i]. \label{Equ:12}
\end{equation}
Differentiating the position with respect to time yields the orbital velocity:
\begin{equation}
\dot{x}_{\rm orb}=\frac{n}{1-e\cos E^{*}}[-A\sin E^{*}+F\sqrt{1-e^{2}}\cos E^{*}], \label{Equ:13}
\end{equation}
\begin{equation}
\dot{y}_{\rm orb}=\frac{n}{1-e\cos E^{*}}[-B\sin E^{*}+G\sqrt{1-e^{2}}\cos E^{*}], \label{Equ:14}
\end{equation}
with $n=2\pi/P_{\rm orb}=K_{1}\sqrt{1-e^{2}}/(a_{1}\sin i)$, where $K_{1}$ is the semi-amplitude of the RV variation (see Table~\ref{Table3}).

The difference in proper motion between two epochs, $t1$ and $t2$, arises primarily from changes in orbital motion:
\begin{equation}
\Delta \dot{x}_{\rm mod} = \dot{x}_{\rm orb}(t1)-\dot{x}_{\rm orb}(t2), \label{Equ:15}
\end{equation}
\begin{equation}
\Delta \dot{y}_{\rm mod} = \dot{y}_{\rm orb}(t1)-\dot{y}_{\rm orb}(t2). \label{Equ:16}
\end{equation}
The O--C analysis has provided some orbital parameters, leaving $\Omega$ and $i$ as unknowns, with ranges $\Omega$ $\in$ [0,360$^{\circ}$] and $i$ $\in$ [0,180$^{\circ}$]. We perform a grid search over this parameter space using a step size of 0.1$^{\circ}$. For each ($\Omega$, $i$) pair, we compute the predicted velocity differences between epochs 2015.5 and 2016.0 and compare them with the observed proper motion anomaly. The optimal parameters are determined by minimizing the following objective function:
\begin{equation}
\chi^{2} = [\frac{\Delta \dot{x}_{\rm mod}(\Omega, i)-\Delta pmDE_{t1-t2}}{\Delta pmDE_{t1-t2}}]^{2} +[\frac{\Delta \dot{y}_{\rm mod}(\Omega, i)-\Delta pmRA_{t1-t2}}{\Delta pmRA_{t1-t2}}]^{2}. \label{Equ:17}
\end{equation}
in which $\Delta$pmDE$_{t1-t2}$ = 0.7305 km s$^{-1}$ and $\Delta$pmRA$_{t1-t2}$ = -0.4229 km s$^{-1}$ represent the proper motion differences between Gaia DR2 and DR3 \citep{2018A&A...616A...1G,2023A&A...674A...1G}.
We used the parallax, Plx, from Gaia DR3 \citep{2023A&A...674A...1G} after applying the systematic zero-point correction recommended by \citet{2021A&A...649A...4L}. Applying the correction yields a corrected parallax of 0.2756 $\pm$ 0.0243 mas for IY Lyr. This corrected value was used to convert proper motions to RV via the relation $Vr$ = 4.74 pm/Plx.

\begin{figure*}
\centering
%\vbox to4.0in{\rule{0pt}{5.5in}}
%\special{psfile=Figure03.eps angle=0 hoffset=0 voffset=0 vscale=100
%hscale=100}
\includegraphics[width=0.45\textwidth]{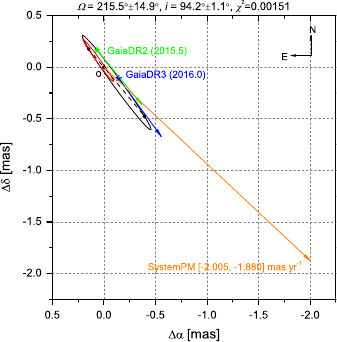}
\caption{Projected orbits of the IY Lyr binary system on the celestial plane for $\Omega$ = 215.5$^{\circ}$ $\pm$ 14.9$^{\circ}$, and $i$ = 94.2$^{\circ}$ $\pm$ 1.1$^{\circ}$. The black curve shows the orbit of the pulsating primary star, and the red curve shows the orbit of the unseen compact companion. The green and blue arrows indicate the orbital velocities (in mas yr$^{-1}$) at the Gaia DR2 and Gaia DR3 epochs, respectively. The orange arrow represents the systemic proper motion of the binary system.} \label{Fig.4}
\end{figure*}

The objective function yields an optimal solution: $\Omega$ = 215.5$\pm$14.9$^\circ$, $i$ = 94.2$\pm$1.1$^\circ$ (errors estimated via Monte Carlo simulation). The grid search also reveals a secondary local solution: $\Omega$ = 84.4$\pm$16.0$^\circ$, $i$ = 85.8$\pm$1.1$^\circ$. This secondary solution is confidently excluded based on LAMOST RV data. Figure~\ref{Fig.4} shows the celestial projection of the orbits for the target star (black curve) and its companion (red curve). The green and blue arrows indicate the orbital velocities (in mas yr$^{-1}$) at the Gaia DR2 and Gaia DR3 epochs, respectively. The orange arrow represents the systemic proper motion of the binary system.

\subsubsection{Extending the Proper Motion Anomaly Method via the Photocentric Weight Factor $q'$}

It is important to note that proper motion data trace the motion of the system's photocenter. If the companion's luminosity is negligible (e.g., a compact object or exoplanet), the photocenter's semi-major axis equals that of the primary star, $a_{\rm photo} = a_{1}$, which is the assumption in the calculation above. For a companion with nonnegligible luminosity, $a_{\rm photo} = q'a_{1}$ with $q'$ $<$ 1, where
\begin{equation}
q'=1-\frac{m_{1}+m_{2}}{m_{2}}\frac{L_{2}}{L_{1}+L_{2}}. \label{Equ:18}
\end{equation}
In the equation, $m_1$ and $m_2$ are the masses of the primary and companion star, respectively, while $L_1$ and $L_2$ denote their luminosities. This relationship can be derived from the formulation provided by \citet{1967pras.book.....V} in terms of the magnitude difference between the two components, and is equivalent to that presented in \citet{2024ApJS..271...50F}.

\begin{figure*}
\centering
%\vbox to4.0in{\rule{0pt}{5.5in}}
%\special{psfile=Figure03.eps angle=0 hoffset=0 voffset=0 vscale=100
%hscale=100}
\includegraphics[width=0.45\textwidth]{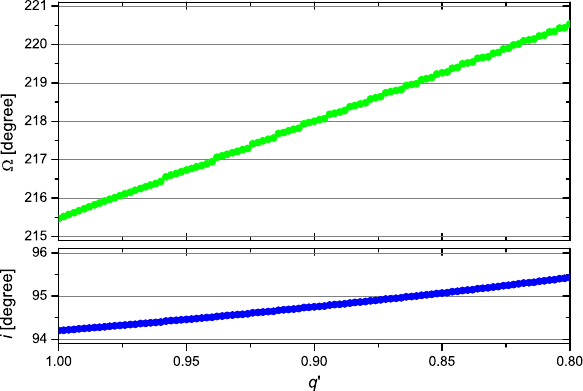}
\caption{Variation of the orbital parameters $\Omega$ (longitude of the ascending node) and $i$ (inclination) as a function of the photocentric weight $q'$. Only modest changes are seen, confirming the robustness of the orbital solution against the unknown luminosity of the companion.} \label{Fig.5}
\end{figure*}

For IY Lyr, the O--C analysis suggests a lower limit for the companion's mass of approximately 1.4\ $M_{\odot}$ (see Section \ref{Sec:Discuss02}). If the companion is a dark compact object (e.g., a massive white dwarf near the Chandrasekhar limit or a neutron star), then $q'$ = 1. If the companion is a main-sequence star, the mass lower limit mentioned above would correspond to an F5 dwarf, whose luminosity is $\sim$ 5 - 6\ $L_{\odot}$ according to the mass-luminosity relation. Considering the mass of the pulsating primary ($\sim$ 0.61 M$_\odot$, see Section \ref{Sec:Discuss01}) and its typical luminosity ($\sim$ 50 $L_{\odot}$), Equation (\ref{Equ:18}) gives $q'$ $\approx$ 0.85, i.e., a fractional light contribution of the companion of $\sim$ 10\%. We systematically explored the parameter space for $q'$ ranging from 1.0 to 0.8. For each value of $q'$, we repeated the grid search to obtain the corresponding ($\Omega$, $i$). The results, presented in Figure~\ref{Fig.5}, show that as $q'$ decreases from 1.0 to 0.85, $\Omega$ increases from 215.5$^{\circ}$ to 219.3$^{\circ}$, and $i$ increases from 94.2$^{\circ}$ to 95.1$^{\circ}$. Although these shifts are statistically significant, their amplitudes are modest, indicating that the primary star dominates the system's light and that the orbital parameters are not highly sensitive to the companion's luminosity. However, based on the analysis of the system's nature presented in Section \ref{Sec:Discuss02}, a main-sequence companion can likely be excluded. Therefore, we adopt the solution for $q' = 1$, which yields $i$ = 94.2$^{\circ}$. This inclination is used in all subsequent calculations.

It should be noted, however, that IY Lyr was not observed by the Hipparcos satellite. Consequently, we cannot utilize the classical proper motion anomaly approach, which leverages the $\sim$ 25 yr time baseline between Hipparcos and Gaia to amplify the orbital motion signal. As an alternative, we employed the proper motion difference between Gaia DR2 and Gaia DR3; however, the time span between these two epochs is only about 0.5 yr. Given the system's orbital period of 3.94 yr, this relatively short baseline may weaken the orbital signal in the proper motion difference, thereby reducing the precision of the derived $i$ and $\Omega$.

Looking ahead, the forthcoming Gaia DR4 will provide time series astrometric data rather than epoch-averaged proper motions. This advancement will enable us to directly model the orbital motion of the IY Lyr system through a joint fit, effectively overcoming the current limitation imposed by a short baseline (e.g., the result of \citealt{2024A&A...686L...2G}). With Gaia DR4, we will significantly improve the accuracy of the orbital parameters and independently verify the conclusions drawn from the proper motion difference analysis presented here. This development offers a promising pathway toward ultimately determining the nature of the companion and its formation channel.

\subsection{Kinematic Parameters and Galactic Population Classification} \label{Sec:PopulationClassification}

To determine the kinematic properties and population membership of the IY Lyr system within the Galaxy, we combined data from various sources for a comprehensive kinematic analysis. The core observational parameters are listed in Table~\ref{Table5}. Positional information was obtained from the CDS website\footnote{\url{http://simbad.cds.unistra.fr/simbad/}}, while RV and proper motion were derived from Sections \ref{sec:RadialVelocity} and \ref{Sec:ProperMotions}, respectively. The distance was calculated using Gaia DR3 parallax data. We employed the galpy package\footnote{\url{https://github.com/jobovy/galpy}} to compute the orbital parameters of the IY Lyr system within the Galactic potential \citep{2015ApJS..216...29B}. The calculations utilized the Milky Way potential model from \citet{2017MNRAS.465...76M}, which includes multiple mass components such as a dark matter halo, stellar disk, and Galactic bulge, providing an accurate description of the Galactic gravitational potential distribution. The adopted solar position and kinematic parameters for the calculations were: solar Galactocentric distance R$_{0}$ = 8.20 kpc, circular velocity at the solar position $v_{0}$ = 232.8 km s$^{-1}$, and solar motion relative to the Local Standard of Rest ($U_{\odot}$, $V_{\odot}$, $W_{\odot}$) = (11.1, 12.24, 7.25) km s$^{-1}$ \citep{2010MNRAS.403.1829S,2020MNRAS.492.2161Z}. By performing a numerical integration of the orbit over 5 Gyr with a time step of 0.5 Myr, we obtained the system's complete orbital parameters. The resulting dynamical parameters are listed in Table~\ref{Table5}, including angular momentum components, total energy, spatial positions and velocity components, as well as characteristic parameters derived from the orbital integration. Observational uncertainties in distance, proper motions, and RV were propagated to all quantities in Table~\ref{Table5} using Monte Carlo simulations (1000 realizations per star), assuming Gaussian errors and the McMillan17 Galactic potential. The reported uncertainties are the standard deviations of the resulting distributions.

The results presented in Table~\ref{Table5} show that the system's Z-component angular momentum ($L_{\rm z}$ = 1287 $\pm$ 35 kpc km s$^{-1}$) exhibits clear prograde motion. Its value lies between the high angular momentum typical of thin disk stars and the near-zero values characteristic of halo stars, falling within the typical range for thick-disk stars (800 - 1800 kpc km s$^{-1}$). Orbital integration indicates that its maximum height above the Galactic plane ($z_{\rm max}$ = 1.17 $\pm$ 0.10 kpc) is significantly greater than that of the thin disk (typically $<$ 0.5 kpc) but lower than that of typical halo stars ($>$ 3 kpc). Its low orbital eccentricity (= 0.258 $\pm$ 0.020) further supports its classification as a disk star. In velocity space, both the system's tangential velocity ($V_{\phi}$ = 181.5 $\pm$ 4.4 km s$^{-1}$) and peculiar velocity with respect to the Local Standard of Rest ($V_{\rm pec}$ = 76.6 $\pm$ 6.2 km s$^{-1}$) fall within the typical dispersion range for thick-disk stars, well below the threshold commonly used to distinguish halo stars \citep{2014A&A...562A..71B,2020MNRAS.492.2161Z}. Crucially, its chemical signature of metal-poor but $\alpha$-enhanced composition ([Fe/H] $\simeq$ -1.0 $\pm$ 0.1, [$\alpha$/Fe] $\simeq$ +0.27 $\pm$ 0.03; \citealt{2019ApJS..245...34X,2024ApJS..272...31W}), aligns consistently with the kinematic characteristics of moderate prograde angular momentum and high-velocity dispersion observed here, forming a synergistic combination distinctive of the thick-disk population.

Based on the age-metallicity relation established by \citet{2022Natur.603..599X} from precise subgiant measurements, the IY Lyr system is a likely member of the old, high-$\alpha$ thick-disc population that formed at $\sim$ 13 Gyr. Its high [$\alpha$/Fe] and low [Fe/H] place it at the metal-poor end of the early disc's age-metallicity sequence.\footnote{The term "metal-poor" here is used in the context of typical Galactic stellar population studies; in the field of RR Lyrae stars, this metallicity is not considered within the metal-poor regime.} This indicates formation during the initial chemical enrichment phase when only core-collapse supernovae had contributed. The high angular momentum, $L_{\rm z}$ = 1287 kpc km s$^{-1}$ demonstrates that this star has remained on a disk-like orbit, escaping significant kinematic heating during the Gaia-Sausage-Enceladus merger approximately 11 Gyr ago. Therefore, it represents a surviving relic of the primordial Milky Way disk that was not "splashed" into the halo \citep{2020MNRAS.494.3880B}.

\section{Discussions}  \label{Sec:Discussion}

\subsection{Mass and period evolution of the pulsating primary star} \label{Sec:Discuss01}
\begin{figure*}
\centering
%\vbox to4.0in{\rule{0pt}{5.5in}}
%\special{psfile=Figure03.eps angle=0 hoffset=0 voffset=0 vscale=100
%hscale=100}
\includegraphics[width=0.45\textwidth]{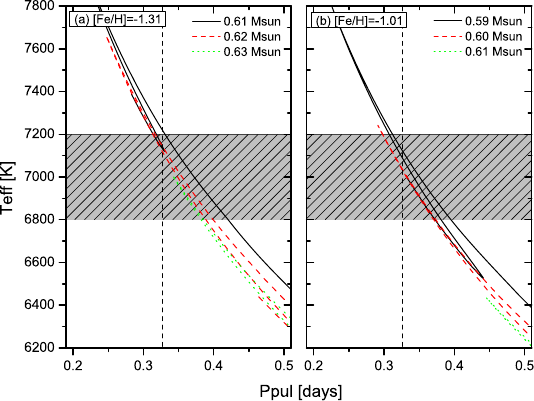}
\caption{Period--temperature diagrams for IY Lyr. Evolutionary tracks from BaSTI models for [Fe/H] = -1.31 and -1.01 are shown. The observed period (0.3266 days) and temperature range (6800--7200 K) intersect the tracks at masses of 0.61--0.62 M$_{\odot}$ ([Fe/H] = -1.31) and 0.59--0.60 M$_{\odot}$ ([Fe/H] = -1.01), yielding a primary mass of $\sim$ 0.61 $M_{\odot}$.} \label{Fig.6}
\end{figure*}

To determine the mass of the RRc star IY Lyr, we utilize BaSTI $\alpha$-enhanced canonical evolutionary models, considering two metallicity cases: $Z$ = 0.002 ([Fe/H] = -1.31) and $Z$ = 0.004 ([Fe/H] = -1.01) \citep{2006ApJ...642..797P}. Based on the horizontal-branch evolutionary parameters, we compute the first-overtone pulsation period using the equation provided by \citet{2015ApJ...808...50M}. The LAMOST LRS spectrum was obtained at the pulsation phase of approximately 0.4, corresponding to a temperature that represents the mean temperature over the entire pulsation cycle. Multiple studies have estimated this temperature to lie in the range of 6800 - 7200 K \citep{2018yCat.5153....0L,2019yCat.5164....0L,2022yCat.5156....0L,2019ApJS..245...34X,2023ApJS..266...40W,2024ApJS..272...31W,2020ApJS..246....9Z,2022ApJS..258...26Z,2024ApJS..271...58D,2026yCat.5162....0L} , which we adopt as the mean (intrinsic) temperature of IY Lyr. In the period--temperature diagram (see Figure~\ref{Fig.6}), the evolutionary tracks for masses of 0.61 and 0.62 $M_{\odot}$ at $Z$ = 0.002, and for 0.59 and 0.60 $M_{\odot}$ at $Z$ = 0.004, intersect the observed pulsation period (0.3266 day) within the temperature range of IY Lyr (see Figure~\ref{Fig.6}). Averaging these values yields a pulsating primary star mass of approximately 0.61 $M_{\odot}$, which we adopt as the basis for further discussion.

Based on the BaSTI evolutionary tracks corresponding to the mass and metallicity of IY Lyr, the predicted secular period change rate for a horizontal-branch RRc star is typically on the order of $\sim$ 0.01 day Myr$^{-1}$. This value is approximately two orders of magnitude smaller in absolute terms than the observed parabolic term, $\beta$ = $-1.24\pm0.14$ day Myr$^{-1}$, derived from the O--C analysis. Therefore, the parabolic trend cannot be attributed to a genuine evolutionary period decrease. Instead, it more plausibly represents a long-segment, nonperiodic irregular variation commonly observed in some RRc stars \citep{2013JAVSO..41...75P}, which can be empirically modeled by a quadratic term. Importantly, the periodic component of the O--C residuals, which we interpret as the LTTE caused by a binary companion, is independently supported by RV and astrometric proper motion analyses (Sections \ref{sec:RadialVelocity} and \ref{Sec:ProperMotions}). Hence, while the parabolic term reflects irregular pulsation behavior, the cyclic signal provides evidence for binarity.

\subsection{Nature of the companion} \label{Sec:Discuss02}

Based on the orbital inclination and the mass of the pulsating primary, the mass function indicates a companion mass of 1.37$\pm$0.19 $M_{\odot}$. The system's age likely exceeds 10 Gyr, ruling out the possibility that the companion is an F-type main-sequence star. Binary evolution models for RR Lyrae systems also exclude the scenario of an accreting main-sequence companion, as such systems would evolve into RR Lyrae variables earlier ($<$ 9 Gyr) due to the larger mass of the progenitor of the pulsating star \citep{2024MNRAS.52712196B}. Therefore, the pulsating primary likely formed through single-star evolution, while the more massive companion is expected to be a compact object resulting from more advanced evolutionary stages.

The mass of 1.37 $M_{\odot}$ lies at the intersection between the upper mass limit for white dwarfs (Chandrasekhar limit of $\sim$ 1.44 $M_{\odot}$) and the peak of the typical neutron star mass distribution ($\sim$ 1.35 $M_{\odot}$), allowing the companion to be interpreted either as a massive O-Ne-Mg white dwarf near the mass limit or as a typical neutron star. If the companion is a white dwarf, its mass would exceed the natural upper limit predicted by single-star evolution ($\sim$ 1.1 $M_{\odot}$) \citep{2000A&A...363..647W}, necessitating mass growth via accretion through binary interactions. However, the IY Lyr system has a relatively long orbital period, corresponding to a separation of about 2--5 astronomical units. The current primary is a 0.61 $M_{\odot}$ RRc star, and its main-sequence progenitor had a mass of only about 1.0 $M_{\odot}$, which could not provide sufficient stellar wind or stable Roche-lobe overflow to enable effective accretion and achieve a mass increase of approximately 0.2--0.3 $M_{\odot}$. Even during the primary's red giant phase, wind accretion would be subject to considerable uncertainties in efficiency and total accreted mass. Moreover, from a statistical perspective, forming a white dwarf with a mass so close to the Chandrasekhar limit requires finely tuned initial conditions and evolutionary parameters, making such objects rare in observed samples.

In contrast, interpreting the companion as a neutron star does not require additional assumptions about accretion-driven growth. A mass of 1.37 $M_{\odot}$ lies well within the core range of the neutron star mass distribution and can be naturally produced via core-collapse supernova from a progenitor with an initial mass of approximately 8--12 $M_{\odot}$ \citep{2025NatAs...9..552Y}. Although supernova explosions typically impart significant kick velocities to newly formed neutron stars \citep{1994Natur.369..127L,2001ApJ...549.1111L,2005MNRAS.360..974H,2008AIPC..983..433K}, which could disrupt wide binaries, systems with low kick velocities or more compact initial orbits may survive to the present day. Considering measurement uncertainties, the likelihood that the companion's mass falls below the typical lower limit for neutron stars ($\sim$ 1.1 $M_{\odot}$) is small, while values near or exceeding 1.44 $M_{\odot}$ remain fully consistent with the neutron star interpretation. In summary, in the absence of direct evidence such as pulsed signals or accretion-powered X-rays, identifying the companion of IY Lyr as a typical neutron star represents the most parsimonious interpretation that is globally consistent with the observational data. Future improvements in mass precision through more accurate photometric and RV measurements, or the detection of radio pulsations or X-ray emission from the companion, will be crucial for confirming its true nature.

\subsection{Comparison with the Results of \citet{2021ApJ...915...50H}} \label{Sec:Discuss03}

\citet{2021ApJ...915...50H} presented a statistical analysis of RR Lyrae binary candidates based on OGLE data toward the Galactic bulge, finding that the mass function distribution exhibits a trimodal shape, corresponding to companion masses of approximately 0.6, 0.2, and 0.067 $M_{\odot}$, respectively. Among their sample, the star OGLE-BLG-RRLYR-20376 (hereafter OGLE-20376) shows a high degree of similarity in orbital parameters to IY Lyr, as presented in this work ($P_{\rm orb}$ = 1641 days, $a_{1}\sin i$ = 2.330 au, $e$ = 0.628, $f(m)$ = 0.66507 $M_{\odot}$). Both stars have very similar orbital periods, projected semi-major axes, and mass functions. Their mass functions are significantly higher than the highest peak of the trimodal distribution reported in \citet{2021ApJ...915...50H}, implying a minimum companion mass of about 1.4 $M_{\odot}$ and suggesting the possibility of a massive white dwarf or a neutron star companion.

Despite their similar orbital parameters, IY Lyr and OGLE-20376 differ in their Galactic population membership. Although OGLE-20376 is located in the direction of the bulge, our dynamical analysis using Gaia data and the RV from \citet{2024A&A...690L..17L} shows that its orbital characteristics are consistent with the Galactic halo population. Its orbit is highly eccentric ($\sim$ 0.82), typical of halo stars with radial, deeply penetrating orbits; the maximum height above the Galactic plane reaches $Z_{\rm max}$ = 4.17 kpc, far exceeding the thick-disk scale height and consistent with halo membership. The angular momentum about the Galactic $Z$-axis is very low ($L_{z}$ = 314.4 kpc km s$^{-1}$), indicating negligible net rotation around the Galactic center. Additionally, its peculiar velocity relative to the Local Standard of Rest is extremely high ($V_{\rm pec}$ = 319 km s$^{-1}$), well above the typical dispersion of disk stars (see Table \ref{Table5}). Furthermore, literature reports extremely low photometric metallicities for this star, down to [Fe/H] = -2.52 \citep{2024A&A...690L..17L} and [Fe/H] = -3.212 \citep{2022ApJS..261...33D}, supporting its classification as a very old halo star. This indicates that the two stars belong to the thick disk and the halo, respectively, representing different star formation and evolutionary histories.\footnote{It should be noted that \citet{2025ApJ...984...58H} derived a metallicity of [Fe/H] = -0.83 for OGLE-20376 using the deep learning model Gaia Net on Gaia DR3 low-resolution spectra. If this value is accurate, combined with its low angular momentum, $L_{\rm z}$ = 314 kpc km s$^{-1}$, the star could be a "splashed" thick-disk star resulting from the merger with the Gaia-Sausage-Enceladus satellite galaxy \citep{2020MNRAS.494.3880B,2022Natur.603..599X}.}

Overall, the bulge sample of \citet{2021ApJ...915...50H} is characterized by a predominance of low-mass companions (white dwarfs, red dwarfs, and brown dwarfs), with orbital periods primarily concentrated between 3000 and 4000 days and an eccentricity peak near 0.25--0.3. In contrast, IY Lyr (a thick-disk RRc star) and OGLE-20376 (a halo RRab star) exhibit similar orbital parameters but have mass functions significantly higher than the typical values observed in the bulge sample, despite belonging to different Galactic substructures. The rarity of systems with such high-mass functions in the bulge sample may reflect their intrinsic scarcity within the bulge. Variations among Galactic substructures in star formation rate, initial mass function, metallicity evolution, and binary interaction history likely contribute to the diversity in formation efficiency and the distribution of orbital parameters in compact companion binary systems \citep{2024MNRAS.52712196B}.

It is important to emphasize that comparisons based on only one or a few high-mass function systems are insufficient to draw universal conclusions. The orbital similarity between IY Lyr and OGLE-20376 may be coincidental, or it may indicate a specific formation channel (e.g., survival of a primordial massive binary after a supernova explosion) that operates efficiently in different environments. A systematic expansion of RR Lyrae binary candidate samples in the thick-disk, halo, and bulge, combined with high-precision RV and metallicity measurements, will be essential to distinguish between these possibilities and to gain a deeper understanding of the formation and evolution of RR Lyrae binaries across different Galactic substructures.

\section{Summary}  \label{Sec:Conclusion}

In this study, we conducted a comprehensive analysis of the pulsation characteristics, binary orbital motion, and Galactic population membership of the RRc star IY Lyr. Our investigation utilized multiband time series photometry (ASAS-SN, ZTF, TESS, and our own $BVRI$ observations), LAMOST spectroscopy, and Gaia astrometry. By cross-validating the O--C method, radial velocity residuals, and proper motion anomaly analysis, we reveal for the first time that this star likely hosts a compact companion. Furthermore, we precisely constrain its orbital parameters, physical properties, and stellar population affiliation. The main conclusions are as follows:

1. \textit{Detection of binary orbital motion.} The O--C diagram reveals a long-term period decrease at a rate of $\beta$ = -1.24 $\pm$ 0.14 day Myr$^{-1}$, superimposed by a LTTE with an orbital period of $P_{\rm orb}$ = 1438.5 $\pm$ 31.4 days ($\approx$ 3.94 yr), eccentricity $e$ = 0.46 $\pm$ 0.15, and mass function $f(m)$ = 0.65 $\pm$ 0.14 $M_{\odot}$. %The eccentric orbit model is significantly preferred over a circular one (p = 0.0011p=0.0011).

2. \textit{Independent verification of the companion and determination of its nature.} The LAMOST RVs, after subtracting the pulsational contribution, agree well with the orbital velocity curve predicted by the O--C solution. Using the proper motion difference between Gaia DR2 and DR3, together with the O--C parameters, we derive an orbital inclination $i$ = 94.2$^{\circ}$ $\pm$ 1.1$^{\circ}$, leading to a companion mass $M_{2}$ = 1.37 $\pm$ 0.19 M$_{\odot}$. The system's age likely exceeds 10 Gyr, ruling out a main-sequence companion. This mass lies at the peak of the neutron star mass distribution and does not require accretion growth, making a typical neutron star the most plausible interpretation; however, a massive white dwarf near the Chandrasekhar limit cannot be completely excluded.

3. \textit{IY Lyr belongs to the old, thick-disk population of the Milky Way.} Dynamical integration yields an angular momentum $L_{\rm z}$ = 1287 $\pm$ 35 kpc km s$^{-1}$, a maximum height above the Galactic plane of $z_{\rm max}$ = 1.17 $\pm$ 0.10 kpc, and an orbital eccentricity of $e$ = 0.258 $\pm$ 0.020. Combined with its metallicity ([Fe/H] $\simeq$ -1.0 $\pm$ 0.1) and $\alpha$ enhancement ([$\alpha$/Fe] $\simeq$ +0.27 $\pm$ 0.03), IY Lyr is identified as an old ($\sim$ 13 Gyr), high-$\alpha$ thick-disk star, providing observational support for binary evolution channels that produce RRLs.

In summary, IY Lyr is likely one of the very few RRc stars with a compact companion ($\simeq$ 1.4 $M_{\odot}$) that has been independently supported through O--C, RV, and astrometric methods. It represents the first clear case of such a system in the thick disk. This work also extends the classical proper motion anomaly method by introducing the photocentric weight factor $q'$, making it applicable to systems where the companion contributes a nonnegligible amount of light. Future Gaia DR4 time series astrometry will enable further refinement of the orbital parameters and ultimately help confirm the true nature of the companion. Our finding carries important implications for thick-disk binary evolution and the formation channels of neutron stars in binary systems.

\acknowledgments

We thank the anonymous referee for their valuable comments and suggestions, which have helped improve the rigor and quality of this work. This work is supported by the International Cooperation Projects of the National Key R\&D Program (No. 2022YFE0127300), the International Partnership Program of Chinese Academy of Sciences (No. 020GJHZ2023030GC), the Yunnan Fundamental Research Projects (grant Nos. 202503AP140013, 202501AS070055, 202401AS070046), the China Manned Space Program with grant No. CMS-CSST-2025-A16, the CAS "Light of West China" Program and "Yunnan Revitalization Talent Support Program" in Yunnan Province. We acknowledge the support of the staff of the ZEISS-600 telescope at the Maidanak Astronomical Observatory and the Xinglong 2.16 m/85 cm telescope. This work was partially supported by the National Astronomical Observatories, Chinese Academy of Sciences.

Based on observations obtained with the Samuel Oschin Telescope 48-inch and the 60-inch Telescope at the Palomar Observatory as part of the Zwicky Transient Facility project. ZTF is supported by the National Science Foundation under Grants No. AST-1440341 and AST-2034437 and a collaboration including current partners Caltech, IPAC, the Oskar Klein Center at Stockholm University, the University of Maryland, University of California, Berkeley, the University of Wisconsin at Milwaukee, University of Warwick, Ruhr University, Cornell University, Northwestern University and Drexel University. Operations are conducted by COO, IPAC, and UW. This paper includes data collected with the TESS mission, obtained from the MAST data archive at the Space Telescope Science Institute (STScI). Funding for the TESS mission is provided by the NASA Explorer Program. STScI is operated by the Association of Universities for Research in Astronomy, Inc., under NASA contract NAS 5-26555. The specific observations analyzed can be accessed via \citet{https://doi.org/10.17909/3y7c-wa45}. This work has made use of data from the European Space Agency (ESA) mission {\it Gaia} (\url{https://www.cosmos.esa.int/gaia}), processed by the {\it Gaia} Data Processing and Analysis Consortium (DPAC, \url{https://www.cosmos.esa.int/web/gaia/dpac/consortium}). Funding for the DPAC has been provided by national institutions, in particular the institutions participating in the {\it Gaia} Multilateral Agreement. This work also utilizes photometric data from the ASAS-SN database \citep{2014ApJ...788...48S,2018MNRAS.477.3145J,2019MNRAS.485..961J,2019MNRAS.486.1907J,2020MNRAS.491...13J,2021MNRAS.503..200J,2023MNRAS.519.5271C}.

\bibliographystyle{aasjournal}
\bibliography{Ref}
\clearpage
%%%%%%%%%%%%%%%%%%%%%%%%%%%%%%%%%%%%%%%%%%%%%%%%%%

\begin{table}[htbp]
\centering
\caption{Summary of Photometric Data Used in This Work for IY~Lyr.}\label{Table1}
\label{tab:photometric_data_summary}
\begin{tabular}{ccccc}
\toprule
Source & Time Range & Band/ & Data Points & Typical Error  \\
                & (HJD-2400000)  & Filter & ($N$)         & (mag)  \\
\hline
ASAS-SN & 56595--58414 & $V$ & 310 & 0.0241  \\
ZTF     & 58204--60610 & $zg$ & 1205 & 0.0111 \\
        & 58197--60610 & $zr$ & 1432 & 0.0101 \\
        & 58229--60576 & $zi$ & 176  & 0.0112 \\
TESS    & 56303--57799 & TESS & 32113 & 0.0050  \\
Maidanak60 & 60189--60568 & $B$ & 670 & 0.0057 \\
             & 60189--60568 & $V$ & 674 & 0.0049  \\
             & 60189--60568 & $R_c$ & 670 & 0.0053 \\
             & 60189--60568 & $I_c$ & 671 & 0.0064 \\
XL85         & 61004--61004 & $V$   &  25 & 0.0050 \\
             & 61004--61004 & $R$   &  25 & 0.0050 \\
\hline
\end{tabular}

\medskip
\footnotesize{
\textbf{Notes.} ASAS-SN: All-Sky Automated Survey for Supernovae \citep{2014ApJ...788...48S}. ZTF: Zwicky Transient Facility \citep{2019PASP..131a8003M}.\\
TESS: Transiting Exoplanet Survey Satellite \citep{2015JATIS...1a4003R}. Its overall operating band range is approxiamately 600--1000 nanometers. Maidanak60: 60 cm telescope at Maidanak Astronomical Observatory. XL85: 85 cm telescope at XingLong Observatory.}
\end{table}

\clearpage
\begin{table}\tiny
\caption{The 180 new available times of light maximum for IY Lyr obtained from the sky surveys, literature and our observations.}\label{Table2}
\begin{center}
\begin{tabular}{ccc|ccc|ccc|ccc}
\hline\hline
 HJD       &   Error  &   Ref.  &    HJD     &  Error & Ref. & HJD     &   Error  &   Ref. &  HJD &  Error  &  Ref. \\
 2400000+   &   (day) &         &   2400000+  & (day) &      & 2400000+ &   (day) &        & 2400000+ & (day) &   \\
\hline
56788.9325 & 0.0062 & ASASSN & 59057.9497 & 0.0037 & ZTF(zg) & 60139.3681 & 0.0100 & ZTF(zr) & 60332.7391 & 0.0025 & TESS \\
56846.4140 & 0.0058 & ASASSN & 59060.2461 & 0.0032 & ZTF(zr) & 60155.7163 & 0.0024 & ZTF(zg) & 60334.0371 & 0.0023 & TESS \\
56906.1868 & 0.0079 & ASASSN & 59115.7698 & 0.0097 & ZTF(zg) & 60189.3574 & 0.0012 & 60cm(R) & 60335.0212 & 0.0066 & TESS \\
57115.1968 & 0.0065 & ASASSN & 59135.0401 & 0.0034 & ZTF(zr) & 60189.3577 & 0.0013 & 60cm(B) & 60335.9992 & 0.0016 & TESS \\
57184.7848 & 0.0086 & ASASSN & 59299.3331 & 0.0065 & ZTF(zr) & 60189.3591 & 0.0016 & 60cm(V) & 60336.9762 & 0.0023 & TESS \\
57253.3847 & 0.0047 & ASASSN & 59353.5499 & 0.0076 & ZTF(zi) & 60189.3592 & 0.0022 & 60cm(I) & 60337.9642 & 0.0023 & TESS \\
57469.5681 & 0.0085 & ASASSN & 59388.8266 & 0.0065 & ZTF(zr) & 60192.2749 & 0.0011 & 60cm(B) & 60338.9385 & 0.0023 & TESS \\
57552.8761 & 0.0179 & ASASSN & 59392.0902 & 0.0011 & TESS   & 60192.2777 & 0.0011 & 60cm(V) & 60425.8078 & 0.0037 & ZTF(zr) \\
57608.4084 & 0.0063 & ASASSN & 59394.3700 & 0.0046 & ZTF(zg) & 60192.2843 & 0.0019 & 60cm(I) & 60425.8103 & 0.0031 & ZTF(zg) \\
57847.1722 & 0.0037 & ASASSN & 59395.0304 & 0.0018 & TESS   & 60192.2854 & 0.0019 & 60cm(R) & 60480.3611 & 0.0026 & TESS \\
57911.5206 & 0.0040 & ASASSN & 59398.3016 & 0.0016 & TESS   & 60195.2282 & 0.0010 & 60cm(V) & 60481.3417 & 0.0021 & TESS \\
57972.2778 & 0.0069 & ASASSN & 59404.5045 & 0.0012 & TESS   & 60195.2290 & 0.0011 & 60cm(B) & 60482.3217 & 0.0017 & TESS \\
58221.5147 & 0.0025 & ZTF(zg) & 59408.0972 & 0.0013 & TESS   & 60195.2294 & 0.0010 & 60cm(R) & 60483.3002 & 0.0030 & TESS \\
58233.5818 & 0.0105 & ASASSN & 59411.0388 & 0.0027 & TESS   & 60195.2409 & 0.0018 & 60cm(I) & 60486.5873 & 0.0084 & TESS \\
58242.4179 & 0.0019 & ZTF(zr) & 59413.9792 & 0.0015 & TESS   & 60196.2130 & 0.0012 & 60cm(V) & 60488.5452 & 0.0034 & TESS \\
58262.9897 & 0.0034 & ZTF(zg) & 59415.2792 & 0.0117 & ZTF(zi) & 60196.2139 & 0.0011 & 60cm(B) & 60489.5141 & 0.0033 & TESS \\
58270.8151 & 0.0066 & ZTF(zi) & 59416.9206 & 0.0013 & TESS   & 60196.2143 & 0.0010 & 60cm(R) & 60490.4966 & 0.0045 & TESS \\
58288.4477 & 0.0047 & ASASSN & 59472.1042 & 0.0066 & ZTF(zi) & 60196.2180 & 0.0022 & 60cm(I) & 60491.4685 & 0.0060 & TESS \\
58304.7949 & 0.0031 & ZTF(zg) & 59478.6353 & 0.0054 & ZTF(zr) & 60197.1923 & 0.0019 & 60cm(R) & 60492.4508 & 0.0042 & TESS \\
58305.4462 & 0.0028 & ZTF(zr) & 59481.5972 & 0.0080 & ZTF(zg) & 60197.1929 & 0.0020 & 60cm(B) & 60493.7588 & 0.0030 & TESS \\
58314.9056 & 0.0073 & ZTF(zi) & 59699.4427 & 0.0173 & ZTF(zr) & 60197.1933 & 0.0015 & 60cm(V) & 60494.7346 & 0.0024 & TESS \\
58338.7575 & 0.0050 & ZTF(zr) & 59707.2788 & 0.0045 & ZTF(zg) & 60197.1976 & 0.0026 & 60cm(I) & 60495.7119 & 0.0043 & ZTF(zg) \\
58345.9390 & 0.0055 & ZTF(zg) & 59729.8277 & 0.0078 & ZTF(zi) & 60209.2673 & 0.0012 & 60cm(R) & 60495.7137 & 0.0021 & TESS \\
58358.9968 & 0.0067 & ZTF(zi) & 59748.4468 & 0.0067 & TESS   & 60209.2729 & 0.0030 & 60cm(I) & 60496.6960 & 0.0035 & TESS \\
58360.3025 & 0.0018 & ASASSN & 59751.3972 & 0.0086 & TESS   & 60209.2749 & 0.0011 & 60cm(V) & 60497.6791 & 0.0020 & TESS \\
58394.6124 & 0.0025 & ZTF(zr) & 59754.3223 & 0.0016 & TESS   & 60209.2752 & 0.0016 & 60cm(B) & 60498.6531 & 0.0020 & TESS \\
58401.7930 & 0.0028 & ZTF(zg) & 59761.5104 & 0.0059 & TESS   & 60217.1122 & 0.0085 & ZTF(zr) & 60499.9618 & 0.0026 & TESS \\
58588.6053 & 0.0038 & ZTF(zg) & 59764.4473 & 0.0096 & TESS   & 60225.6125 & 0.0049 & ZTF(zg) & 60500.9435 & 0.0019 & TESS \\
58598.0771 & 0.0038 & ZTF(zr) & 59767.3896 & 0.0018 & TESS   & 60313.4641 & 0.0046 & TESS   & 60501.9251 & 0.0023 & TESS \\
58647.3958 & 0.0015 & ZTF(zg) & 59771.3073 & 0.0013 & TESS   & 60314.4416 & 0.0032 & TESS   & 60502.9067 & 0.0024 & TESS \\
58661.7684 & 0.0024 & ZTF(zr) & 59774.2440 & 0.0016 & TESS   & 60315.4184 & 0.0042 & TESS   & 60502.9085 & 0.0045 & ZTF(zr) \\
58664.0522 & 0.0054 & ZTF(zi) & 59777.5190 & 0.0066 & TESS   & 60316.4008 & 0.0026 & TESS   & 60503.8910 & 0.0058 & TESS \\
58685.6041 & 0.0031 & ZTF(zg) & 59780.4503 & 0.0013 & TESS   & 60317.3841 & 0.0028 & TESS   & 60504.8620 & 0.0040 & TESS \\
58721.2129 & 0.0036 & ZTF(zg) & 59781.4272 & 0.0059 & ZTF(zg) & 60319.3444 & 0.0033 & TESS   & 60514.3362 & 0.0011 & 60cm(V) \\
58739.8317 & 0.0051 & ZTF(zr) & 59784.0435 & 0.0013 & TESS   & 60320.6488 & 0.0018 & TESS   & 60514.3368 & 0.0023 & 60cm(R) \\
58767.5974 & 0.0042 & ZTF(zg) & 59787.6399 & 0.0014 & TESS   & 60321.6304 & 0.0021 & TESS   & 60514.3379 & 0.0011 & 60cm(B) \\
58957.3588 & 0.0041 & ZTF(zg) & 59788.9330 & 0.0083 & ZTF(zr) & 60322.6122 & 0.0020 & TESS   & 60514.3443 & 0.0021 & 60cm(I) \\
58984.7841 & 0.0043 & ZTF(zr) & 59790.2427 & 0.0056 & ZTF(zi) & 60323.5893 & 0.0022 & TESS   & 60556.7925 & 0.0103 & ZTF(zg) \\
59011.9059 & 0.0017 & TESS   & 59790.5827 & 0.0054 & TESS   & 60324.5687 & 0.0017 & TESS   & 60568.2298 & 0.0015 & 60cm(B) \\
59014.8424 & 0.0022 & TESS   & 59846.7522 & 0.0123 & ZTF(zi) & 60325.5446 & 0.0024 & TESS   & 60568.2307 & 0.0015 & 60cm(V) \\
59017.7836 & 0.0016 & TESS   & 59860.7974 & 0.0040 & ZTF(zr) & 60327.8393 & 0.0056 & TESS   & 60568.2356 & 0.0017 & 60cm(R) \\
59021.3802 & 0.0026 & TESS   & 59860.8033 & 0.0024 & ZTF(zg) & 60328.8122 & 0.0027 & TESS   & 60568.2420 & 0.0028 & 60cm(I) \\
59024.6441 & 0.0017 & TESS   & 60067.5226 & 0.0061 & ZTF(zr) & 60329.7895 & 0.0024 & TESS   & 60573.7784 & 0.0058 & ZTF(zr) \\
59027.5848 & 0.0045 & TESS   & 60076.3574 & 0.0035 & ZTF(zi) & 60330.7674 & 0.0024 & TESS   & 61004.9324 & 0.0014 & 85cm(V) \\
59030.8468 & 0.0012 & TESS   & 60084.5075 & 0.0033 & ZTF(zg) & 60331.7571 & 0.0022 & TESS   & 61004.9336 & 0.0014 & 85cm(R) \\
\hline\hline
\end{tabular}
\end{center}
\footnotesize{
\textbf{Notes.} 60cm: 60 cm telescope at Maidanak Astronomical Observatory. 85cm: 85 cm telescope at XingLong Observatory.}
\end{table}

\begin{table}
\begin{center}
\caption{The pulsating and orbital elements of IY Lyr. $P_{\rm orb}$ is the orbital period, $T$ is the time of passage through the periastron, $f(M)$ is the mass function of the companion, $K_{1}$ is the velocity semi-amplitude in km s$^{-1}$, and $\chi^{2}$ is the residual sum of squares. $\Omega$ and $i$ are obtained from the proper motion anomaly method (See Section \ref{Sec:ProperMotions}).}\label{Table3}
\begin{tabular}{l c }
\hline\hline
Parameter                &   LTTE fitting             \\
\hline
$T_{0}$[cor]             & 2458291.05922(119)         \\
$P_{\rm pul}$[cor](days) & 0.32662127(36)             \\
$\beta$ (days cycle$^{-1}$) & $-1.11(.12)\times10^{-9}$  \\
$\beta$ (days Myr$^{-1}$)   & $-1.24(.14)$               \\
$a_{1}\sin i/c$ (days)      & 0.01247(89)                \\
$a_{1}\sin i$ (au)       & 2.16(.15)                  \\
$e$                      & 0.46(.15)                  \\
$\omega$ (deg)    & 135.1(18.3)                \\
$P_{\rm orb}$ (days)          & 1438.5(31.4)             \\
$P_{\rm orb}$ (yr)         & 3.938(86)                \\
$T$                      & 2458384.7(72.4)            \\
$f(M)$ ($M_{\odot}$)     & 0.65(.14)                  \\
$K_{1}$ (km s$^{-1}$)        & 18.35(2.08)            \\
$\chi^{2}$               & 1642.754                   \\
\hline
$\Omega$ (deg)    & 215.5(14.9)                \\
$i$ (deg)         & 94.2(1.1)                  \\
\hline\hline
\end{tabular}
\end{center}
\end{table}

\begin{table*}
\begin{center}
\caption{RVs of IY Lyr Corrected for Pulsational Variations Using Templates from \citet{2017ApJ...848...68S,2021ApJ...919...85B,2024RAA....24g5009H}.}\label{Table4}
\begin{tabular}{c c | c c | c c | c c | c c}
\hline\hline
Obs Time &           &  Wang2024 &                  & Sneden2017 &                       & Braga2021  &                       & Huang2024  &  \\
\hline
HJD      & Pulsation & Vr\_metal & Vr\_$\rm \alpha$ & Vr\_m\_cor & Vr\_$\rm \alpha$\_cor & Vr\_m\_cor & Vr\_$\rm \alpha$\_cor & Vr\_m\_cor & Vr\_$\rm \alpha$\_cor \\
2400000+ & Phase     & km s$^{-1}$ & km s$^{-1}$    & km s$^{-1}$ & km s$^{-1}$          & km s$^{-1}$ & km s$^{-1}$          & km s$^{-1}$ & km s$^{-1}$ \\
\hline
LRS         &      &       &       &       &       &       &       &       &  \\
57169.2419  & 0.41 & -38.5(2.9) & -38.5(1.8) & -42.2 & -43.4 & -42.5 & -46.0 & -41.5 & -42.3  \\
\hline
MRS         &      &       &       &       &       &       &       &       &  \\
59016.1797  & 0.10 & -49.4(0.7) & -54.0(1.1) & -37.6 & -33.2 & -37.3 & -34.3 & -35.8 & -36.4  \\
59016.1964  & 0.15 & -45.8(0.6) & -50.4(0.7) & -36.4 & -32.9 & -36.6 & -33.4 & -34.4 & -34.2  \\
59016.2123  & 0.20 & -43.5(0.5) & -46.4(1.3) & -36.9 & -33.0 & -37.1 & -33.2 & -35.2 & -32.8  \\
\hline\hline
\end{tabular}
\end{center}
\end{table*}

\begin{table}
\begin{center}
\tiny
\caption{The Galactic positions and the kinematics of the IY Lyr system and OGLE-BLG-RRLYR-20376 (OGLE-20376) system.}\label{Table5}
\begin{tabular}{l l l l}
\hline\hline
Parameter                &       IY Lyr  &   OGLE-20376   &  Source/Description   \\
Observation Parameters   &               &                &          \\
%\hline
R.A. ($\alpha$, J2000) (deg)  & 277.41987  & 263.23184 &   CDS        \\
Decl. ($\delta$, J2000) (deg)       & 31.00016   & -22.76635 &   CDS        \\
Parallax ($\varpi$) (mas)              & 0.2756(242)      & 0.2495(996)     & GaiaDR3     \\
Distance ($d$) (kpc)                       & 3.628(.319)      & 4.008(1.600)     & (1/$\varpi$) \\
System proper motion in R.A. ($\mu\alpha$*$\cos\delta$) (mas yr$^{-1}$)  &  -2.005(14) & -10.823(.118) & This work/GaiaDR3 \\
System proper motion in Decl. ($\mu\delta$) (mas yr$^{-1}$)              & -1.880(15)  & -5.578(77) & This work/GaiaDR3  \\
Syestm RV ($V_{\rm r}$) (km s$^{-1}$)                      & -37.3(3.0)  & 228.4(35.5) & This work/Luongo et al. (2024) \\
\hline
Current Position and Velocity &     &  &\\
Galactocentric radius ($R$) (kpc)    &   7.09(3)     & 4.24(1.52) &  Projected distance from the Galactic center\\
Height above the Galactic plane ($Z$) (kpc) &  1.13(.10)   & 0.41(.15) &  Current vertical height\\
RV ($V_{\rm R}$) (km s$^{-1}$) &  54.5(4.0)  & -235.6(39.8) & Galactocentric RV    \\
Tangential velocity ($V_{\rm T}$ or $V_{\phi}$) (km s$^{-1}$) &   181.5(4.4)  & 74.2(61.1) & Azimuthal velocity  \\
Vertical velocity ($V_{\rm Z}$) (km s$^{-1}$) &   15.1(1.9)  & 144.9(45.3) &  Velocity perpendicular to the Galactic plane\\
Peculiar velocity ($V_{\rm pec}$) (km s$^{-1}$) &   76.6(6.2) & 319.0(85.8) & Velocity relative to LSR  \\
\hline
angular Momentum \& energy   &   &   & \\
$L_{\rm Z}$ (kpc km s$^{-1}$)   & 1287(35)  & 314(419) & Angular momentum about the Galactic $Z$-axis    \\
$L_{\bot}$ (kpc km s$^{-1}$)    & 210.2(14.5)   & 710.8(104.8) & Angular momentum perpendicular to the $Z$-axis  \\
Total energy ($E_{\rm tot}$) ((km s$^{-1}$)$^{2}$) & -171400(610)  & -176200(17700) & Sum of kinetic and potential energy \\
\hline
Orbital Characteristics   &   &   & \\
Maximum height above the plane ($Z_{\rm max}$) (kpc) & 1.17(.10) & 4.17(.95) & Maximum vertical excursion from orbit integration \\
Pericentrer distance ($R_{\rm peri}$) (kpc) & 4.55(.18) & 0.85(1.02) & Closest approach to the Galactic center \\
Apocenter distance ($R_{\rm apo}$) (kpc) & 7.70(5)  & 8.6(4.0)  & Farthest point from the Galactic center \\
Orbital eccentricity & 0.258(20) & 0.820(.115) &  ($R_{\rm apo}$-$R_{\rm peri}$)/($R_{\rm apo}$+$R_{\rm peri}$) \\
Orbital inclination [$^{\circ}$] & $\simeq$ 9.3(.8) & $\simeq$ 66.1(21.3) & Estimated from $\arctan$ ($L_{\bot}$/$L_{\rm Z}$) \\
\hline\hline
\end{tabular}
\end{center}
\footnotesize{
\textbf{Notes.} IY Lyr uses the parallax corrected for the Gaia zero-point \citep{2021A&A...649A...4L} for consistency with its proper motion analysis; OGLE-20376 uses the uncorrected value, since the correction is smaller than its statistical error.}
\end{table}

%
%
%\clearpage
%%%%%%%%%%%%%%%%% APPENDICES %%%%%%%%%%%%%%%%%%%%%
%\appendix



\end{document}